# ReFilter: Bridging Embeddings and LLM Filtering for Similar Mobile App Retrieval

**AUTHORS SECTION**

**AlMulla, Buthayna** University of Toronto, Canada | buthayna.almulla@mail.utoronto.ca

**Assi, Maram** Université du Québec à Montréal, Canada | assi.maram@uqam.ca

**Hassan, Safwat** University of Toronto, Canada | safwat.hassan@utoronto.ca

## ABSTRACT

Retrieving similar mobile applications (apps) is essential for researchers, developers, and end-users. Researchers use similarity detection to study app ecosystems and trends, developers for competitor analysis, and end-users for focused app recommendations. Existing approaches rely on embedding-based retrieval, which captures semantic similarity but fails to identify functionally similar apps. To our knowledge, no prior work has applied large language model (LLM)-based filtering to this task, due to the high computational cost of evaluating large numbers of app pairs. To address this gap, we propose **ReFilter**, a hybrid framework that first **Retrieves** semantically related candidate apps using embeddings and then applies LLM-based contextual **Filtering** to identify true functionally similar apps with higher precision. This design balances efficiency and accuracy, achieving an F1-score of 90% for retrieving similar apps. By improving the relevance of app alternatives, ReFilter enables more accurate app comparisons and supports improved ecosystem understanding, competitor analysis, and recommendations.

## INTRODUCTION

App stores are among the largest publicly accessible information environments, hosting millions of applications (apps) with rich metadata, including descriptions, ratings, categories, app screenshots, and user reviews (Harman et al., 2012; Martin et al., 2017). This metadata provides valuable insights into how information systems evolve, compete, and overlap in functionality. Prior studies leveraged app store data to investigate system evolution, release behavior, user feedback, and peer-app competition (McIlroy et al., 2016; Hassan et al., 2022; Guzman & Maalej, 2014), demonstrating the analytical potential of these platforms. In particular, Al-Subaihin et al. (2021) showed that developers actively monitor similar apps for feature discovery, idea validation, and system evolution, highlighting app stores as a key source of competitive and functional insights. The structure and visibility mechanisms of app stores position them as large-scale information environments that influence how developers access, interpret, and use information about competing similar apps. However, accurately identifying functionally similar apps remains challenging, as apps may share features, categories, or audiences without serving as true functional substitutes.

We define functionally similar apps as *apps that users are likely to select as substitutes for a given target* app (Uddin et al., 2020). This definition emphasizes functional relevance, focusing on whether apps satisfy the same user need rather than merely sharing textual, semantic, or market similarity. Accurately retrieving functionally similar apps supports key information tasks, including comparative information seeking for competitive analysis, feature discovery for requirements elicitation, and information organization through categorization and clustering.

Recent work on app similarity has primarily relied on embedding-based retrieval, where semantic vector representations of app descriptions or metadata are used to identify related apps at scale (Alsubaihin et al., 2019). While these models effectively capture textual and semantic similarity, they often fail to represent deeper functional relationships (Linares-Vásquez et al., 2016). Two apps may share similar textual semantics yet target different user needs. For example, a Spanish-learning app and a Danish-learning app are semantically related language-learning applications, but they are not functionally similar because they satisfy different user needs.

Contextual understanding is equally important in identifying similar apps. Large Language Models (LLMs) excel at interpreting such context, reasoning about an app's purpose, prioritizing core functionality, and eliminating noisy features with an intent beyond surface semantics (Zhu et al., 2025). However, their use in large-scale retrieval pipelines is computationally expensive and impractical for thousands or millions of app pairs (Yang et al., 2024).

To overcome this trade-off between scalability and contextual accuracy, we propose **ReFilter**: a hybrid framework that combines embedding-based Retrieval with LLM-based contextual Filtering. Our framework operates in two stages: (1) an embedding stage that retrieves a semantically close candidate set for each target app, and (2) a filtering stage where an LLM refines this set through a functional relevance assessment. This design leverages embeddings for efficiency and LLMs for precision, producing a scalable yet context-aware system for similar app retrieval.

To our knowledge, this is the first study to apply context-aware LLM filtering to large-scale mobile app retrieval. We not only evaluate the effectiveness of different retrieval strategies but also examine how app similarity is represented and varies across functional domains and market scales. We structure our investigation as a three-stage analysis, where each research question builds on the previous one to progressively improve and interpret how functionally relevant app alternatives are identified and accessed.

- **RQ1:** How effectively can embedding-based retrieval capture similar candidate apps?
- **RQ2:** How effectively can LLMs filter retrieved candidates to identify functionally similar apps?
- **RQ3:** What patterns emerge in the identified similar apps across categories?

Across the three research questions, our findings reveal a clear progression in retrieval quality and insight. In **RQ1,** we evaluate 10 target apps against 20,635 apps, and find that text-only embeddings outperform multimodal ones in capturing functional similarity, and that LLM-based embedding models achieve higher F1-Scores (60% at $k=45$) than BERT-based encoders. In **RQ2**, we show that integrating LLM filtering into the retrieval process improves the F1-score, reaching 90% without fixing the number of retrieved apps. Finally, in **RQ3**, we apply our refined framework to a statistically representative sample of 1,156 apps and observe strong variation in the number of retrieved similar apps across categories and download ranges, demonstrating the importance of a dynamic retrieval threshold rather than a static top-*k* cutoff. Together, these findings highlight the complementary strengths of embeddings for scalable candidate retrieval and LLMs for contextual filtering, demonstrating how combining both enables more precise identification of functionally similar apps. This paper makes the following contributions:

1. We propose **ReFilter**, a hybrid **information retrieval framework** that combines embedding-based candidate selection with LLM-based contextual **filtering**, balancing efficiency and accuracy in identifying functionally similar apps.
2. We provide an **empirical evaluation of app similarity retrieval**, demonstrating how embedding choice and LLM-based filtering improve the precision of identifying functionally similar apps.
3. We provide **empirical insights into the structure of app ecosystems**, showing that the number of functionally similar apps varies across categories, download counts, and temporally, motivating the use of dynamic retrieval strategies rather than fixed *top-k* cutoffs.
4. We release a **replication package** that includes scripts from our analysis to support reproducibility. This package enables developers and researchers to reproduce our framework and apply it for identifying functionally similar apps. The replication package is available at: https://github.com/b-almulla/ReFilter

## RELATED WORK

Research on app similarity spans multiple methodological paradigms, from early text-based clustering to modern neural and multimodal retrieval frameworks. Existing approaches can be broadly grouped into four categories: (1) conventional textual similarity methods, which rely on handcrafted or statistical features (Al-Subaihin et al., 2019; Al-Subaihin et al., 2016; Uddin et al., 2021); (2) embedding-based approaches, which generate dense semantic representations from textual descriptions; (3) multimodal techniques, which incorporate additional artifacts such as permissions, reviews, or app screenshots; and (4) code-based approaches, which analyze source code or binary artifacts to infer implementation similarity (Al-Subaihin et al., 2016; Bogomolov et al., 2020; Hamednai et al., 2019; Linares-Vásquez et al., 2016; McMillan et al., 2012; Sanz et al., 2012; Sun et al., 2020).

In this work, we build on **embedding-based representations**, as they align with our goal of identifying functionally similar apps using publicly accessible app store data such as descriptions. Prior work shows that embeddings outperform **conventional textual similarity methods**, which lack contextual and semantic depth. While **multimodal approaches** incorporate additional artifacts such as images, permissions, or reviews, they often capture descriptive or behavioral similarity rather than true functional equivalence. In this work, we specifically evaluate whether incorporating image-based representations (i.e., app screenshots) improves the identification of functionally similar apps, but find limited benefit. **Code-based approaches** capture implementation-level similarity but do not reflect how functionality is presented to users. As our focus is on identifying functionally similar apps from an end-user and ecosystem perspective, we prioritize representations derived from app store data.

### Embedding-Based App Similarity (Text-Only Methods)

Several works focus on leveraging traditional embeddings for this task by encoding each app's description into a vector representation and computing semantic proximity. For instance, prior research developed an embedding-based search engine for Google Play apps that encodes app descriptions into vector representations using the BAAI General Embedding (BGE) model (Wei et al., 2024). Their goal was to retrieve semantically similar features rather than similar apps, but their method provides a strong baseline for semantic retrieval. We adopt BGE as our baseline because it represents a Bidirectional Encoder Representations from Transformers (BERT)-like embedding model commonly used in this domain.

Subsequent work has extended this approach through summarization and clustering. For instance, some studies used generative AI or feature extraction (e.g., NER with BERT, LDA, and NMF) to summarize app descriptions before embedding and clustering them (Alam et al., 2024; Saeed et al., 2024). While such summarization improves conciseness, it often omits details critical for functional comparison, resulting in feature-level rather than holistic app-level similarity. Despite these advances, purely text-based approaches share a common limitation: they capture surface-level or semantic similarity rather than deeper contextual or functional relationships. Two apps may describe overlapping features yet differ significantly in their purpose or target audience. This limitation motivates our exploration of LLM-assisted filtering that can reason about app functionality beyond textual semantics.

### Multimodal App Similarity (Text + Other Artifacts)

To improve upon text-only retrieval, prior work has incorporated diverse artifacts such as permissions, app screenshots, user reviews, market data, and source code metrics to construct richer app representations (Chang et al., 2016; Chen et al., 2015; Nayebi et al., 2016; Uddin et al., 2020; Wei et al., 2022). While these multimodal approaches broaden the feature space, they often conflate *functional similarity* with descriptive, behavioral, or topical similarity derived from app descriptions, permissions, and reviews. In contrast, our work isolates *functional similarity* as the primary competition criterion, focusing on whether apps can serve as true substitutes rather than merely coexisting in the same thematic or market space. Given this focus, we investigate whether app screenshots provide additional evidence of functionality beyond what is described in app descriptions, but find limited benefit.

## METHODOLOGY

We propose **ReFilter**, a hybrid retrieval framework that identifies functionally similar apps by combining the scalability of embedding-based retrieval with the contextual precision of LLMs. Given a *target app* and a corpus of 20,635 apps, the retrieval task is to identify apps that serve as functional substitutes for the target. Throughout this paper, we refer to the app being queried as the *target app*, the retrieved top-k apps as *candidate apps*, and the apps retained after filtering as *similar apps*. ReFilter operates in two stages: **Candidate Retrieval**, which retrieves a top-k set of semantically similar candidate apps using cosine similarity over app embeddings, and **LLM Filtering**, which contextually evaluates each target-candidate pair using a generative LLM to determine whether the candidate serves as a functional substitute for the target. This design balances efficiency and accuracy by leveraging embeddings for scalable retrieval while reserving the LLM for fine-grained reasoning, avoiding costly pairwise comparisons across the full corpus. Figure 1 illustrates the full workflow.

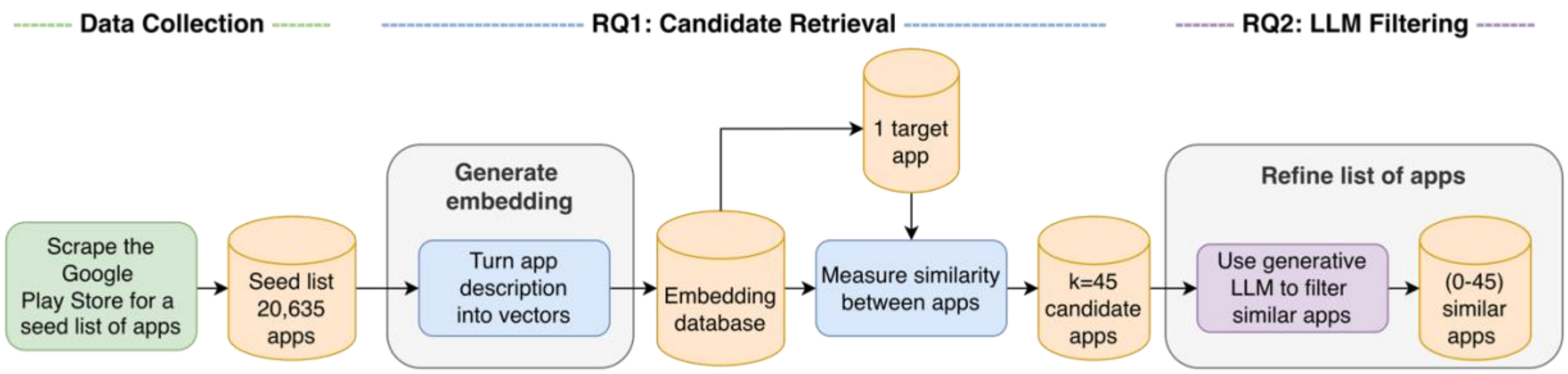

**Figure 1. Framework Outline**

### Data Collection

We collect data by scraping the Google Play Store, retrieving apps from each category page. Following prior work (Uddin et al., 2020; Wei et al., 2024), we exclude the Game category due to its distinct characteristics, as well as the Family category, which primarily contains games (Padla, 2015). We expand coverage by retrieving similar apps listed on the platform and scrape individual app pages to extract metadata, including *app ID, full description, category, downloads, release date, similar apps listed by Google Play, and app screenshots*. Candidate Retrieval and LLM Filtering use only the preprocessed full app description as the textual input; app IDs are used for identification. Categories, download counts, release dates, and Google Play's similar-app lists are not used to determine similarity; they support sampling, RQ3 analysis, and comparison with Google Play. For multimodal embedding methods, we additionally embed the screenshots available on the app page. Data collection was conducted in August 2025, resulting in a seed set of 34,364 apps.

To prepare app descriptions for Candidate Retrieval, we adopt and extend the preprocessing procedure from prior work (Wei et al., 2024). We filter non-English descriptions, remove noisy content (e.g., privacy policies, subscription details, contact information), strip emojis, normalize whitespace, lowercase text, and remove HTML tags. We also exclude apps with descriptions shorter than 200 characters and fewer than 100 downloads to eliminate insufficient and low-quality entries. After preprocessing, 20,635 apps remain.

### Manual Validation and Ground Truth Construction

We construct an **evaluation dataset** by randomly selecting one app from each of the top 10 categories and retrieving the top 50 candidates from each embedding method for manual labeling. To reduce method-specific bias, candidate pairs retrieved by different embedding methods are merged into a single annotation dataset, and each target-candidate pair is evaluated only once. This dataset is used to compare embedding methods and identify the best-performing configuration for candidate retrieval (RQ1). For LLM-based filtering (RQ2), we reuse this dataset as a **development dataset** for prompt engineering, using the selected configuration (*Linq* with *k=45*) to refine the filtering process. To assess generalizability, we construct a separate **test dataset** consisting of 10 target apps drawn from 10 new unseen categories and evaluate performance on unseen data using the finalized prompt.

Each target-candidate app pair is manually evaluated to determine whether the two apps are similar. Following prior work (Uddin et al., 2020), we define similar apps as functional alternatives. Annotators first identify the target app's primary user-facing purpose and then evaluate whether the candidate app could satisfy the same user need. Apps sharing only secondary features are not considered similar; for example, a snoring app and a sleep-remedy app are labelled as dissimilar unless the sleep-remedy app also supports snore recording. Annotation guidelines are available in the replication package. Using this criterion, the first and second authors independently annotate an initial statistically significant sample at the 95% confidence level and a 10% margin of error. We calculate the inter-rater agreement using Cohen's Kappa (Cohen, 1960), which results in *kappa = 0.79,* indicating substantial agreement. The two authors discuss disagreements and refine the annotation guidelines to resolve ambiguities and establish a shared interpretation of functional similarity. The first author labels the remaining annotations. In total, our ground truth consists of 1,315 app pairs in the evaluation dataset, 450 in the development dataset, and 450 in the test dataset.

To measure accuracy for RQ1 and RQ2 we compute precision, recall, and F1-score for each target app, then average these scores across all targets to obtain the overall scores.

**Precision** measures the proportion of retrieved candidate apps that are truly similar to the target app. Let *TP* denote the number of True Positives and *FP* the number of False Positives:

$$Precision = \frac{TP}{TP + FP}$$

**Recall** measures the proportion of relevant apps retained after LLM filtering within the retrieved candidate set. Let $\boldsymbol{TP_{LLM}}$ denote the number of *TP* retained after LLM filtering, and let $\boldsymbol{TP_{method}}$ denote the number of *TP* retrieved by the method prior to filtering:

$$Recall = \frac{TP_{LLM}}{TP_{method}}$$

**Pooled recall** measures the proportion of true similar apps retrieved by a given method relative to the similar apps identified across all methods. This metric captures the coverage of each method within a pooled set of relevant apps and follows standard information retrieval pooling practices (Sanderson, 2010). Let $TP_{method}$ denote the number of *TP* retrieved by a given method, and let $TP_{all\ methods}$ denote the union of *TP* identified across all retrieval methods:

$$Pooled\ Recall = \frac{TP_{method}}{TP_{all\ methods}}$$

**F1-score** represents a balanced measure of filtering performance by combining precision and recall:

$$F1 - Score = \frac{2 \times \text{Precision} \times \text{Recall}}{Precision + Recall}$$

### RQ1: Candidate Retrieval

We design a scalable framework for retrieving a high-quality set of semantically similar candidate apps for each target app. We generate vector embeddings for all apps, encoding each app into a fixed-length representation that enables efficient similarity computation. We compare six embedding methods across two data modalities: text-only models using app descriptions and multimodal models that combine app descriptions with app screenshots from the Google Play Store. For each target app, we compute cosine similarity against all other apps and select the *top-k* most similar apps as candidates. We evaluate multiple values of *k (i.e., 5, 10, …, 50*) across embedding methods to assess how retrieval quality varies with candidate set size. We select the best-performing $k$ = 45 with *Linq* embedding model for LLM filtering. Table 1 summarizes the six evaluated methods.

BGE is a BERT-based model with limited input length; therefore, app descriptions are split into smaller segments or "chunks". We evaluate two strategies for representing long descriptions. (*BGE-AVG*) encodes each chunk separately and aggregates chunk embeddings using average pooling, a common strategy for representing long texts (Dai et al., 2022; Shen et al., 2018). (*BGE-FIRST*) uses only the first chunk, representing a simple truncation-based baseline. The third dataset (Linq) encodes full app descriptions without truncation.

To assess the impact of visual information, we evaluate multimodal embeddings using the open-source SigLIP2 model, which aligns text and app screenshots through contrastive learning and has shown strong performance in image retrieval (Czerwinska et al., 2025). Since SigLIP2 does not produce a single embedding for text and images, we explore aggregation strategies, including average pooling for balanced representations *(SigLIP2-AVG)* and a k-max-inspired method that selects and averages the most similar cross-modal pairs (*SigLIP2-TopN*) (Chen et al., 2021; Shen et al., 2018). We further evaluate a closed-source multimodal model that jointly embeds text and images into a single unified representation, eliminating the need for post-hoc pooling (*Voyage AI*). We use the Voyage AI embedding API, powered by MongoDB, as an upper-bound benchmark to assess the benefits of fully integrated multimodal representations.

| Method Name | Model | Data Modality | Pooling Type |
|---|---|---|---|
| BGE-AVG | BGE-small-en-v1.5[1] | Text Only | Average pooling |
| BGE-FIRST | BGE-small-en-v1.5 | Text Only | First chunk only |
| Linq | Linq-Embed-Mistral[2] | Text Only | None (full text embedding) |
| SigLIP2-AVG | SigLIP2 / so400m-long[3] | Multimodal (Text + Image) | Average pooling |
| SigLIP2-TopN | SigLIP2 / so400m-long | Multimodal (Text + Image) | Top-$K$ similarity aggregation |
| Voyage AI | Voyage-multimodal-3[4] | Multimodal (Text + Image) | None (Unified embedding) |

**Table 1. Summary of Embedding Models, Modalities, Pooling Types**

To measure similarity between apps, we compute pairwise cosine similarity between their embedding vectors. Cosine similarity is widely used in software and app-retrieval research because it effectively captures the angular distance between high-dimensional representations while being invariant to vector magnitude (Alam et al., 2024; Al-Subaihin et al., 2016; Bogomolov et al., 2020; Hamednai et al., 2019; Linares-Vásquez et al., 2016; McMillan et al., 2012; Surian et al., 2017; Wei et al., 2024; Wei et al., 2022). For each target app, we calculate its cosine similarity with every app in the seed list, rank them, and select the top-k as candidates.

### RQ2: LLM Filtering

We refine the candidate apps retrieved in the previous step and identify which ones are *functionally similar* to the target app. While embedding-based retrieval efficiently captures semantic proximity, it lacks contextual understanding of functionality and user intent. To address this limitation, we apply LLM filtering to reason about each candidate's purpose and determine whether it serves as a similar app. For each target app, we feed the list of $k = 45$ candidate apps generated in the previous stage into the LLM, along with their textual descriptions. We evaluate several prompting strategies and few-shot configurations to improve filtering quality.

We select OpenAI's GPT-5[5] for its superior precision in pairwise judgments for LLM filtering. As GPT-5 does not support a temperature of 0, limiting deterministic reproducibility, we run each experiment three times and report average precision, recall, and F1-score to ensure stable performance estimates. The LLM is prompted to evaluate whether each candidate app listed serves as a functional substitute for the target app based on app descriptions, produce a similarity verdict, and justify its decision. This process yields a refined list of similar apps for each target, along with rationales for the analysis. The prompt and few-shot examples are provided in the replication package.

To understand the model's filtering behavior and sources of misclassification, we manually analyze all false negative cases from the zero-shot configuration of the development dataset, which totals 75 instances. For each target-candidate pair, we review the model's justification behind misclassification and categorize the primary cause of confusion into one of five error categories. Based on the error analysis findings, we select few-shot examples that represent recurring filtering challenges observed across multiple target apps rather than isolated cases. These examples illustrate generalizable patterns of confusion that guide the model's contextual filtering.

[1] https://huggingface.co/BAAI/bge-small-en-v1.5
[2] https://huggingface.co/Linq-AI-Research/Linq-Embed-Mistral
[3] https://huggingface.co/fancyfeast/so400m-long
[4] https://voyageai.com
[5] https://developers.openai.com/api/docs/models/gpt-5

## RQ3: Implementing Framework and Analysis

We evaluate the broader applicability of our framework by applying the retrieval and filtering configuration validated in RQ1 and RQ2 to a large-scale dataset of 1,156 apps. We first generate a random set of target apps drawn from the fifteen categories with the largest number of apps. For each category, we calculate a representative sample size using a 95% confidence level and a 10% margin of error, ensuring statistical coverage across domains. This process yields a total of 1,156 target apps, each treated as an independent retrieval query. We then apply ReFilter to identify functionally similar apps for each target. Candidate retrieval incurs a one-time embedding generation cost, whereas GPT-5 filtering introduces per-query inference costs. Across the 1,156 target apps analyzed in RQ3, GPT-5 filtering required approximately 4.2 seconds and $0.04 USD per target app query on average.

To assess whether the number of retrieved similar apps varies across download ranges, we apply the Kruskal-Wallis test (Kruska1 & Wallis, 1952). When significant differences are detected, we perform post-hoc comparisons using Dunn's test with Holm correction (Dunn, 1964). To compare the number of similar apps retrieved by our framework against Google Play's "Similar Apps" lists, we use the Wilcoxon signed-rank test (Wilcoxon, 1992). We apply K-Means clustering (Warren Liao, 2005) to group apps based on temporal similarity trends, representing each app as a yearly time series of similar app counts, a method widely used for pattern discovery in structured data (Ikotun et al., 2023). To determine the optimal number of clusters, we compute silhouette scores (Ashari et al., 2023) across multiple values of *k* and select *k=3*.

## FINDINGS

In RQ1, we evaluate six embedding-based retrieval methods across text-only and multimodal configurations. *Linq* achieves the highest F1-score 60% at *k=45*, showing that LLM-based embeddings capture functional similarity more effectively than smaller or multimodal models. In RQ2, we build on embedding-based retrieval by applying LLM filtering to the candidate sets to improve the accuracy of the final results. Across the test dataset, LLM filtering improves retrieval quality over embeddings alone, with the 2-shot configuration achieving the highest overall F1-score of 90%. In RQ3, we apply the framework to a representative sample of 1,156 target apps and find that the number of similar apps varies by category and download counts, highlighting the need for a dynamic top-k.

## RQ1: How effectively can embedding-based retrieval capture similar candidate apps?

This section presents the findings of the RQ1 experiments, which evaluate the effectiveness of different embedding-based retrieval methods for candidate app identification. Specifically, we examine which embedding method achieves the strongest overall performance, whether incorporating multimodal information improves retrieval accuracy, and how retrieval performance varies with different values of *k*. Figure 2 summarizes these results, showing precision, pooled recall, and F1-score as a function of *k* for each method.

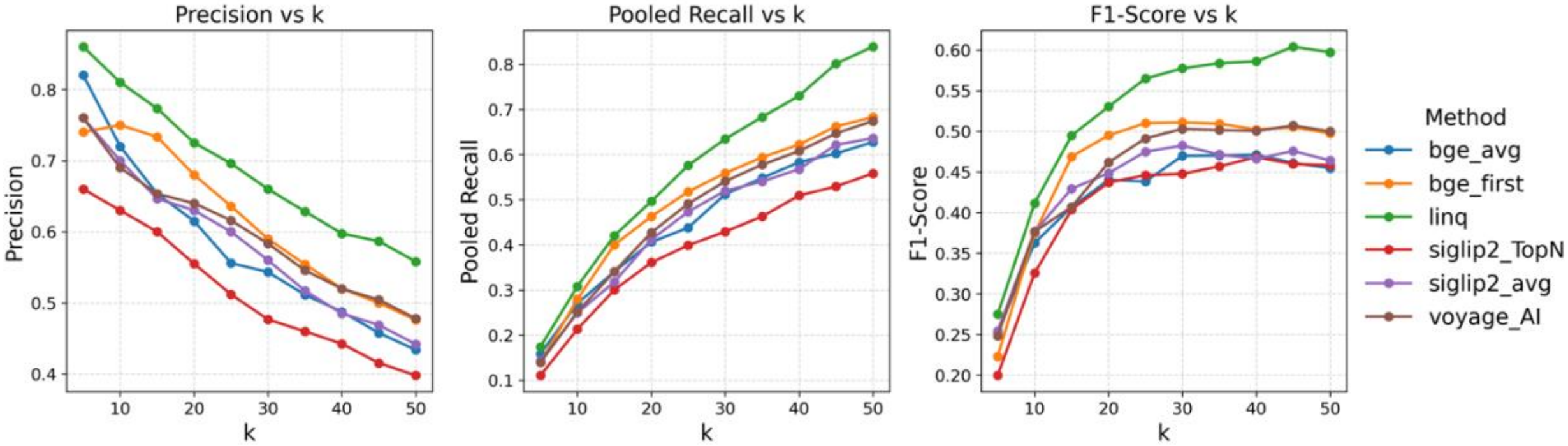

**Figure 2. Precision, pooled recall, F1-score for each method by *k***

### Finding 1.1: Linq model yields the highest precision, pooled recall, and F1-score across all values of *k*

This result suggests that larger LLM-based embedding models, such as *Linq*, capture richer semantic and contextual information than smaller transformer encoders like *BGE*, and that model capacity plays a greater role in retrieval accuracy than pooling strategy. Moreover, *BGE-FIRST* slightly exceeds *BGE-AVG*, implying that key functional cues often appear early in app descriptions.

### Finding 1.2: Multimodal methods offer limited gains

Multimodal models that combine app description and app screenshots, such as *SigLIP2* and *Voyage AI*, do not outperform text-only methods. Screenshots introduce noise (e.g. interface layout) that does not support functional similarity. For example, password managers and expense-tracking apps may present visually similar list-based interfaces despite serving different purposes. For *SigLIP2*, average pooling outperforms the *TopN* variant, indicating that uniformly aggregating text and image embeddings is more effective than emphasizing only the strongest cross-modal matches. *Voyage AI* outperforms *SigLIP2* at k ≥ 20 but shows comparable or lower F1-scores at smaller k.

### Finding 1.3: Embedding-only retrieval remains limited in accuracy

Despite the differences across model architectures and modalities, all embedding-based methods achieve modest F1-scores, with the best performance achieved by *Linq* reaching only 60% at *k=45*. This motivates the next step in our hybrid framework, which is LLM filtering to increase precision, recall, and the overall F1-score.

## RQ2: How effectively can LLMs filter retrieved candidates to identify functionally similar apps?

This section presents the findings of the RQ2 experiments, which examine how LLMs enhance the retrieval accuracy of embedding-based methods. We feed the top *k* = 45 candidate apps retrieved by the *Linq* embeddings into OpenAI's GPT-5 for pairwise evaluation against each target app.

### Finding 2.1: *The 2-shot configuration produces the most effective results, achieving an* F1-score of 90%

As shown in Table 2, introducing 2-shot examples improves the F1-score compared to the 0-shot baseline. While the 0-shot setup already achieves high precision, adding two examples increases recall with minimal impact on precision. Beyond this point, the F1-score declines slightly, suggesting that a small number of examples improves coverage while maintaining labeling accuracy, whereas additional examples provide limited benefit.

| Few-Shot | Precision | Conditional Recall (CR) | (CR) F1-Score | Pooled Recall (PR) | (PR) F1-Score |
|---|---|---|---|---|---|
| 0-shot | 99% | 74% | 83% | 61% | 80% |
| 2-shot | 97% | 84% | 90% | 70% | 86% |
| 4-shot | 98% | 84% | 89% | 70% | 86% |
| 6-shot | 97% | 84% | 89% | 70% | 85% |

**Table 2. Performance Across Few-Shot Configurations on the Test Dataset**

### Finding 2.2: The LLM filtering step improves F1-score compared to embedding-only methods

From RQ1, the best (PR) F1-score achieved by embedding-only retrieval is 60% at *k=45*, while the highest precision reaches 86% at *k=5*. LLM-based filtering improves both metrics, increasing (PR) F1-score to 86% and precision to 97%. Notably, while embedding-based retrieval achieves its highest precision at small *k*, LLM filtering surpasses this performance without relying on a fixed cut-off. A key advantage of LLM filtering is its ability to dynamically determine the number of similar apps per target. Instead of returning a fixed number of candidates, it returns zero apps when no valid alternatives exist and up to 45 when many relevant matches are present. For example, for one app in our test set with no functionally similar alternatives, the model correctly returned zero similar apps, demonstrating robustness in avoiding false positives.

### Finding 2.3: False negatives in the 0-shot setting reveal the model's cautious filtering behavior

Manual analysis of the 75 false negative cases Table 3 shows that most errors occur when the LLM undervalues secondary core features, treats different implementation methods as evidence of dissimilarity, and discounts apps that serve only a part of the core functionality. This pattern suggests that while the model avoids false positives, it is overly conservative when assessing functional overlap. Incorporating few-shot examples that represent these recurring filtering challenges improves recall by helping the model generalize to cases where functionality is shared but presented differently across apps.

| Reason | Cases | Representative Example |
|---|---|---|
| Core feature secondary in candidate app | 28 | *Target:* A password manager. *Candidate:* An authenticator app that includes a password management function. *Rationale*: The LLM dismissed similarity because password management was not the main feature, though users could reasonably use it as an alternative. |
| Same goal, different implementation method | 18 | *Target:* A Spanish learning app. *Candidates:* Other Spanish learning apps using different teaching approaches (e.g., video explanations). *Rationale:* The LLM focused on the learning method rather than the shared goal of language instruction. |

| Reason | Cases | Representative Example |
|---|---|---|
| Candidate app fulfills part of core feature | 11 | *Target:* An all-in-one astrology app; horoscope, tarot, numerology. *Candidates*: Apps focused on either horoscope, tarot, or numerology. *Rationale*: The LLM focused on capturing all core features, when under our heuristics an app capturing at least one could be an alternative. |
| Misidentified core feature | 10 | *Target*: An offline password manager. *Candidate*: A cloud-based password manager. *Rationale*: The LLM overemphasized "offline" functionality, missing that both apps serve the same password management purpose. |
| Different purpose, same core feature | 8 | *Target*: An app designed for mute individuals to communicate via text-to-speech. *Candidates*: General text-to-speech apps. *Rationale*: The LLM judged them dissimilar due to differing purposes, even though both serve the same core communication need for that user. |

**Table 3. Reasons for False Negatives in Zero-Shot LLM Evaluation**

### RQ3: What patterns emerge in the identified similar apps across categories?

This section aims to understand how app similarity manifests across different market segments. By applying our ReFilter framework to a representative sample of 1,156 apps, we analyze how the number of retrieved similar apps varies by category and popularity. This allows us to examine how the availability of functionally similar alternatives differs across app ecosystems. Our analysis highlights the importance of a dynamic retrieval framework that adapts to domain diversity rather than relying on a fixed top-$k$ cutoff.

#### Finding 3.1: The number of similar apps varies across categories

Variation by target category is shown in Figure 3. *Personalization*, *Social*, and *Photography* apps show the highest counts, while categories such as *Business*, *Tools*, and *Sports* have fewer. This pattern indicates that visually or design-driven app categories tend to generate dense clusters of similar apps, whereas functional or utility-oriented categories are more distinct.

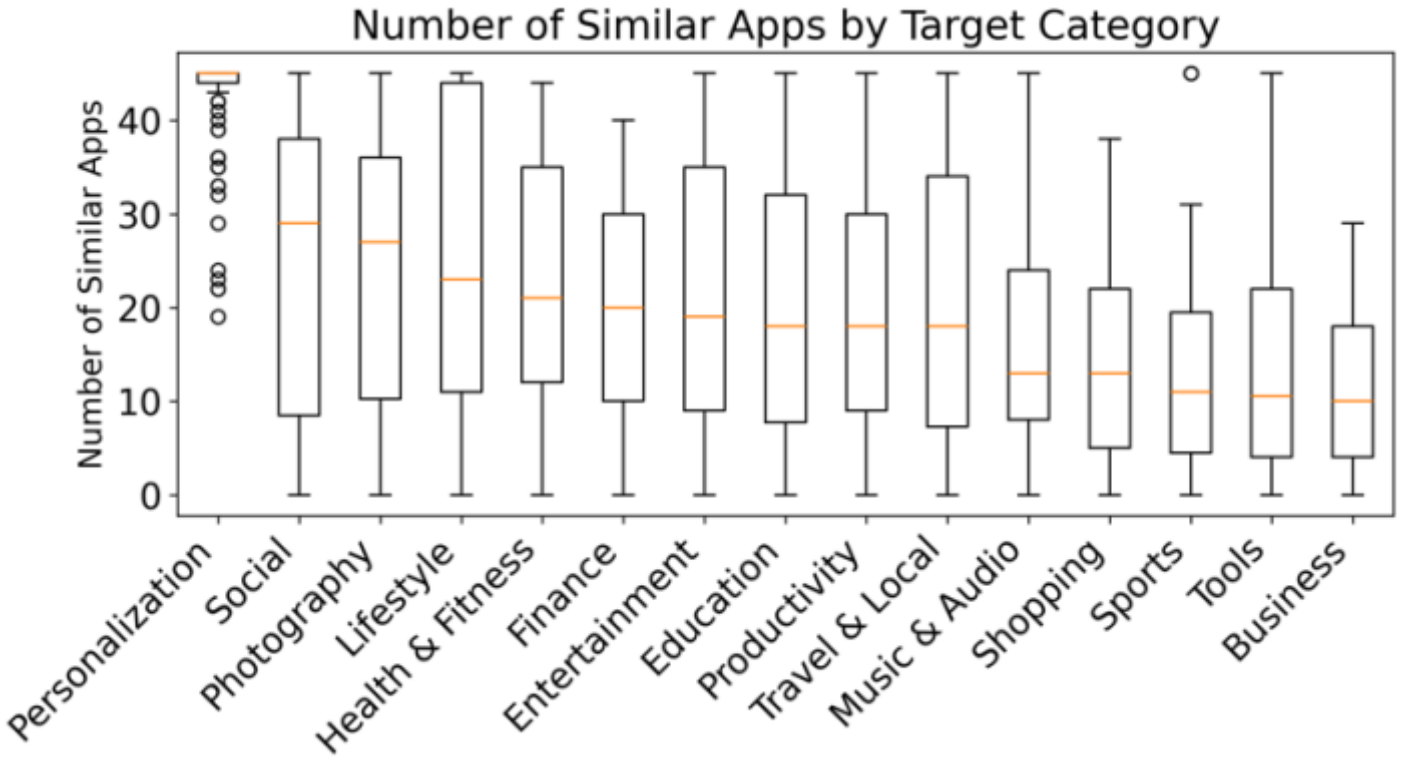

**Figure 3. Number of Similar Apps by Target Category**

#### Finding 3.2: Apps with fewer downloads have significantly more similar apps

Variation by downloads is presented in Figure 4. A Kruskal-Wallis test confirms this effect (H = 180.0, $p < 0.001$, $\epsilon^2$= 0.15). Post-hoc Dunn tests (Holm-corrected) show that apps in the *500+* and *1,000+* ranges have significantly more similar apps than almost all higher ranges ($p < 0.001$ for most comparisons). These low-download outliers are primarily *watch apps*, which appear as clusters of near-identical variants.

#### Finding 3.3: Our method retrieves fewer but more functionally relevant similar apps than Google Play

Across all apps, Google Play consistently returns a significantly larger number of similar apps (mean ≈93) compared to our framework (mean ≈ 21), with this difference confirmed as statistically significant by a Wilcoxon signed-rank test ($p < 0.001$). In most cases (1,031 out of 1,156 apps), Google Play returns more apps, indicating a systematic difference in how similarity is defined.

This difference stems from the design of Google Play's recommendation system, which prioritizes discoverability and user engagement rather than strict functional equivalence. According to Google Play documentation, recommendations aim to surface relevant and appealing apps by leveraging signals such as quality, ratings, reviews, downloads, user appeal, and interaction patterns (Google Play, n.d.-a; Google Play, n.d.-b). As a result, the similar apps feature reflects a discovery-oriented paradigm, promoting exploration and broader exposure rather than identifying direct substitutes (App Radar, 2024). In contrast, our framework explicitly targets functional similarity, identifying apps that users could substitute without losing core functionality. This results in a smaller but more precise set of alternatives, and in some cases uncovers relevant apps not surfaced by Google Play. These findings highlight that while Google Play is effective for recommendation and discovery, it is not designed to capture functional equivalence, highlighting the value of our framework for research and developer-oriented analyses.

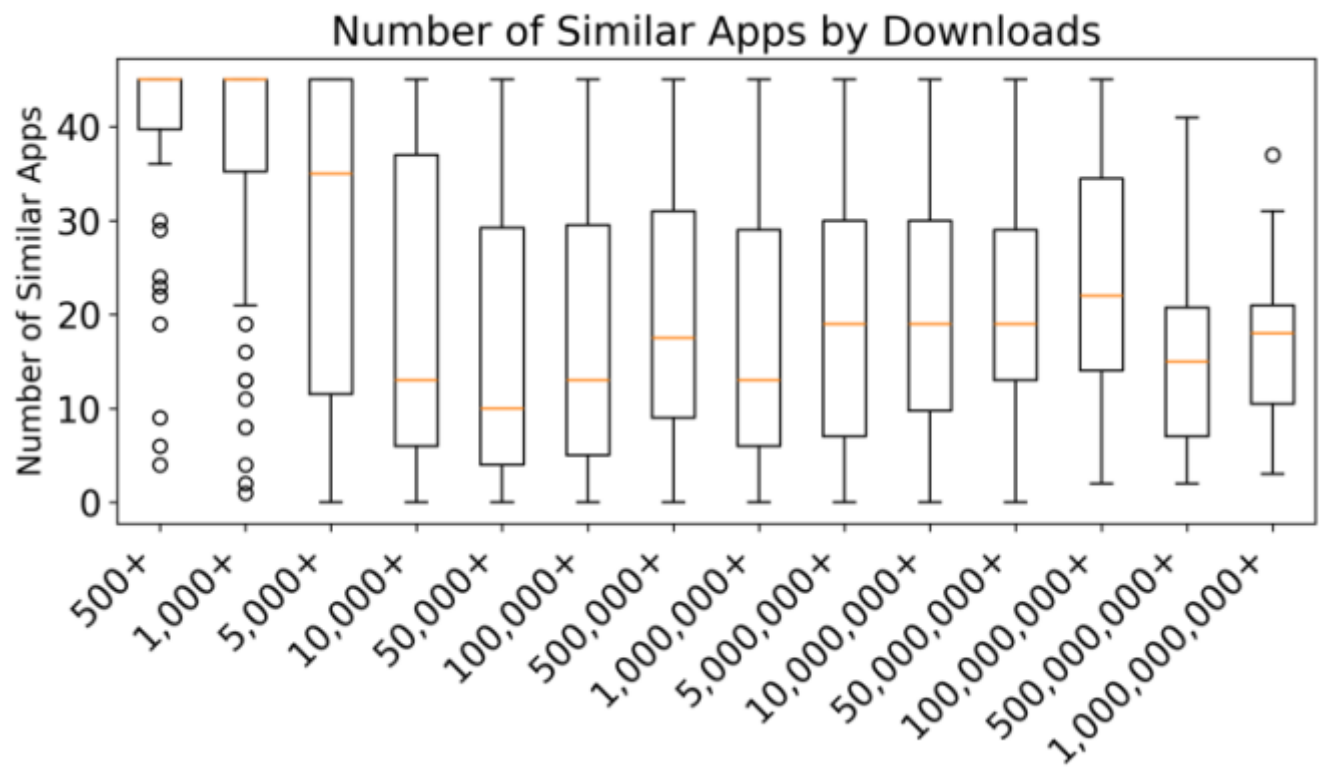

**Figure 4. Number of Similar Apps by Download Range**

### Finding 3.4: Distinct temporal growth patterns reveal different similarity dynamics across apps

Figure 5 presents the clustering of apps based on the temporal evolution of the number of similar apps retrieved from our framework. We identify three distinct growth patterns: Slow Linear Growth, Early Rapid Growth, and Late Rapid Growth. These patterns suggest that the timing of app release and the maturity of the app domain play a key role in shaping how quickly similar alternatives emerge.

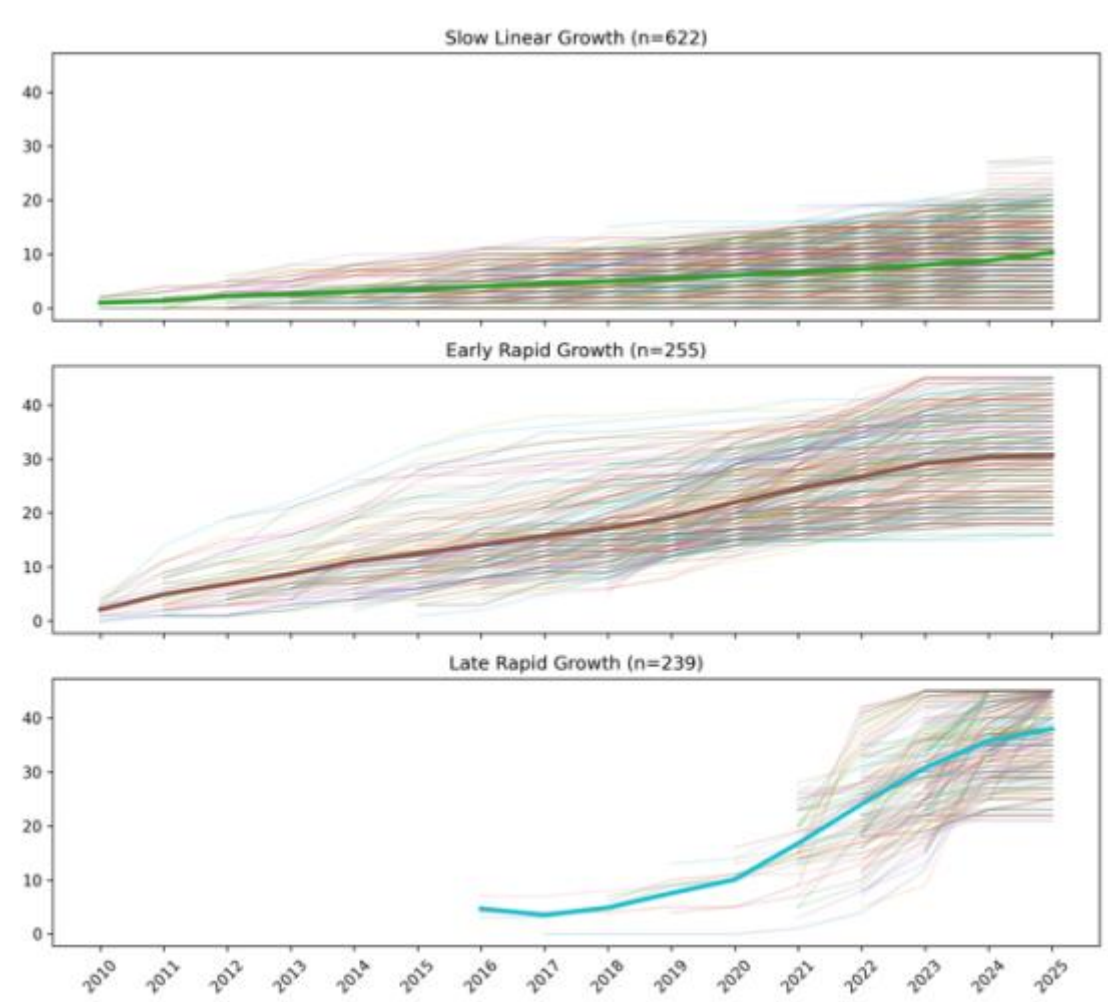

**Figure 5. Time-series Clustering of Apps based on the Number of Similar Apps Over Time**

The *Slow Linear Growth* cluster exhibits a steady and gradual increase in the number of similar apps over time. As shown in the figure, these apps accumulate competitors slowly, without sharp inflection points. Supporting this, the cluster is highly diverse (normalized entropy = 0.98) and spans a wide range of categories, indicating that similarity emerges organically across heterogeneous app domains. The average release year (2019) and relatively high average downloads suggest that these apps belong to a broad, mature ecosystem where competition develops incrementally.

The *Early Rapid Growth* cluster shows a sharp increase in similar apps early in the lifecycle, followed by a gradual stabilization. This pattern indicates that these apps quickly enter competitive spaces and reach saturation relatively early. Consistent with this, the cluster contains older apps (average release year = 2016) and maintains high diversity (normalized entropy = 0.97), suggesting that these apps operate in established domains where functional alternatives emerge early and stabilize over time.

In contrast, the *Late Rapid Growth* cluster is characterized by delayed but steep increases in similarity, with most growth occurring after 2020. The figure shows near-zero values for earlier periods, followed by sharp upward trends. This cluster is more homogeneous (entropy = 0.82) and dominated by specific categories such as Personalization, with repeated developers contributing multiple similar apps. The average release year (2022) and lower downloads suggest that these are recent apps in rapidly expanding, often template-driven ecosystems with many similar variants.

Furthermore, the diversity of temporal growth patterns indicates that there is no one-size-fits-all number of similar apps that can be retrieved across all cases. As shown by the three clusters, the number of similar apps varies by app type, release period, and maturity. Some apps accumulate similar alternatives gradually, others reach saturation early, while newer apps may experience rapid expansion in a short time frame. This variation highlights the limitation of fixed *top-k* approaches and reinforces the importance of a dynamic retrieval strategy, where the number of similar apps is determined by functional similarity to the app rather than by a predefined cutoff.

## DISCUSSSION

Our findings carry distinct implications for researchers, app developers and practitioners, and end users. For **researchers**, our results demonstrate that our hybrid retrieval framework can improve the precision of identifying functionally similar apps. This advances research on information retrieval and enables more reliable datasets for empirical studies on competition, feature evolution, and software ecosystems. Our framework advances app store mining beyond surface-level textual similarity. For **app developers and practitioners**, our framework can help them identify functionally similar apps for competitive analysis. Developers may use these insights to benchmark features, explore design alternatives, and detect near-duplicate apps that overlap in functionality or market positioning. For **end users**, our framework offers a path toward functionality-aware recommendation and categorization systems. Unlike popularity-driven similarity lists, a functionally grounded retrieval system could improve app discoverability, enhance precision in search results, and better reflect real alternatives.

## LIMITATIONS

We acknowledge limitations in the form of internal, construct, and external threats to validity**. Internal validity** threats concern factors within the experimental setup that may have influenced our findings. To ensure fair comparison, all embedding models are evaluated under identical retrieval conditions, including the same candidate pool sizes *k[5, 50],* cosine similarity metric, and preprocessing. By systematically varying *k*, we evaluate precision, pooled recall, and F1-score across different retrieval depths, ensuring that performance differences reflect model capabilities rather than sensitivity to a fixed cutoff.

**Construct validity** relates to whether our measurements accurately capture the intended concept of similarity. We define functional similarity as two apps providing substitutable core functionality, following prior work on app competition and feature similarity (Uddin et al., 2020). However, app descriptions (our main textual input) may contain incomplete, falsely advertised, or marketing-oriented content that does not fully capture functionality. To reduce noise, we apply rigorous preprocessing (e.g., removing short descriptions) and use manual validation to ensure the ground truth aligns with functional equivalence. Furthermore, our recall metric is computed relative to a pooled relevance set derived from the union of top-50 candidates retrieved by all methods. This reflects standard IR pooling practice (Sanderson, 2010) but may not capture all truly similar apps in the full corpus; therefore, recall values should be interpreted as lower bounds rather than absolute recall.

**External validity** addresses the generalizability of our results beyond the studied sample. Our dataset is collected from the Google Play Store, and the results may not fully generalize to other app ecosystems such as the Apple App Store. However, by sampling 20,635 apps across 48 diverse categories, including both utility-oriented (e.g., Tools, Productivity) and content-focused domains (e.g., Lifestyle, Music & Audio), we mitigate store-specific bias and ensure broad, representative coverage of app types. To reduce selection bias, we randomly select 20 target apps from 20 Google Play Store categories and manually evaluate 1,765 unique target-candidate app pairs. However, the evaluation is still based on a limited number of target apps, which may affect the generalizability of the reported performance. Future work could expand the evaluation to additional apps and categories.

## CONCLUSION

This paper advances research on app similarity by introducing ReFilter, a hybrid framework that combines embedding-based retrieval with LLM-based contextual filtering. Our results show that this framework improves retrieval effectiveness, achieving up to 90% F1-score. Unlike prior work that relies on fixed top-*k* retrieval, our framework dynamically determines the number of similar apps per target, reflecting the varying density of functional alternatives across app domains and enabling more accurate identification of truly substitutable apps. Beyond performance improvements, our work provides a reproducible framework for studying app similarity, supports developers in more informed competitor analysis and feature discovery, and enables the design of more meaningful and interpretable app recommendation systems.

## GENERATIVE AI USE

We confirm that we did not use generative AI tools/services to author this submission.

## AUTHOR ATTRIBUTION

First Author: conceptualization, data curation, formal analysis, investigation, methodology, software, validation, visualization, writing – original draft, writing – review and editing; Second Author: conceptualization, investigation, methodology, validation, writing – review and editing; Third Author: conceptualization, investigation, methodology, supervision, validation, funding acquisition, writing – review and editing.

## ACKNOWLEDGMENTS

We acknowledge the support of the Natural Sciences and Engineering Research Council of Canada (NSERC), **[RGPIN-2021-03969]**.